\documentclass[amsmath,superscriptaddress,showpacs,aps,prb,twocolumn]{revtex4-1}
\usepackage[colorlinks,linkcolor=blue,anchorcolor=blue,citecolor=blue,urlcolor=blue]{hyperref}
\usepackage{amsmath}
\usepackage{amssymb}
\usepackage{graphicx}
\usepackage{color}
\usepackage{bm}

\begin{document}
\title{Topological magnons in a one-dimensional itinerant flat-band ferromagnet}
\author{Xiao-Fei Su}
\affiliation{National Laboratory of Solid State Microstructures and Department of Physics, Nanjing University, Nanjing 210093, China}
\affiliation{School of Physics and Electronic Information, Huaibei Normal University, Huaibei 235000, China}
\author{Zhao-Long Gu}
\affiliation{National Laboratory of Solid State Microstructures and Department of Physics, Nanjing University, Nanjing 210093, China}
\author{Zhao-Yang Dong}
\affiliation{National Laboratory of Solid State Microstructures and Department of Physics, Nanjing University, Nanjing 210093, China}
\author{Jian-Xin Li}
\email[]{jxli@nju.edu.cn}
\affiliation{National Laboratory of Solid State Microstructures and Department of Physics, Nanjing University, Nanjing 210093, China}
\affiliation{Collaborative Innovation Center of Advanced Microstructures, Nanjing University, Nanjing 210093, China}
\date{\today}

\begin{abstract}

\par Different from previous scenarios that topological magnons emerge in local spin models, we propose an alternative that itinerant electron magnets can host topological magnons. A one-dimensional Tasaki model with a flat band is considered as the prototype. This model can be viewed as a quarter filled periodic Anderson model with impurities located in between and hybridizing with the nearest-neighbor conducting electrons, together with a Hubbard repulsion for these electrons. By increasing the Hubbard interaction, the gap between the acoustic and optical magnons closes and reopens while the Berry phase of the acoustic band changes from 0 to $\pi$, leading to the occurrence of a topological transition. After this transition, there always exist in-gap edge magnonic modes which is consistent with the bulk-edge correspondence. The Hubbard interaction driven transition reveals a new mechanism to realize non-trivial magnon bands.

\end{abstract}

\maketitle

\section{Introduction}
\par Band structure with non-trivial topology\cite{BLD_RMP2016} has been one of the most active fields in condensed matter physics since the discovery of topological insulators\cite{HK_RMP2010,QZ_RMP2011}. Although the pioneering works focused on bands of fermionic quasi-particles\cite{H_PRL1988,K_PU2001,KM_PRL2005,BHZ_S2006,QHRZ_PRL2009,YXL_PRL2011}, the underlying concepts, such as Berry phase or Berry curvature\cite{B_PRSA1984}, Chern number\cite{TKNN_PRL1982,S_PRL1983}, and Dirac or Weyl point\cite{WAVS_PRB2011}, apply equally to systems that host bands of bosonic excitations. Actually, extensive theoretical and experimental works have been devoted to several areas with bosonic elementary excitations, ranging from topological photonics\cite{LJS_NP2014} to topological phononics\cite{PP_PRL2009}.

\par Magnons, the bosonic quanta of collective spin-1 excitations in a system with a magnetically ordered ground state, also exhibit band structures in crystals. Recently, the search of topological magnons has attracted much attention\cite{OIKSNT_S2010,ZRWL_PRB2013,CHFSBNL_PRL2015,LLKBYC_NC2016,MHJM_PRL_2016,O_JPCM2016,RNF_NJP2016,MS_PRB2017,SWW_PRB2017,LF_PRL2017,M_NP2017,MHM_PRB2017,LLHLF_PRL2017,YLWXDIKLFL_arXiv2017,BWWCLMWRDAWYJW_arXiv2017}, not only because of fundamental interest but also due to its possible applications in spintronics\cite{CVSH_NP2015}. It was proposed theoretically\cite{OIKSNT_S2010,ZRWL_PRB2013} that a Kagome lattice ferromagnet, whose Hamiltonian includes the direct Heisenberg exchange as well as Dzyaloshinskii-Moriya (DM) interaction\cite{D_JPCS1958,M_PR1960}, can host magnon bands with non-zero Chern numbers in the framework of linear spin wave theory. In such system, by the standard Holstein-Primakof (HP) transformation\cite{HP_PR1940}, the local spin model can be mapped to a free bosonic model, with the DM term acting as a vector potential for the propagation of magnons similar to the magnetic field for electrons. Thus non-zero Berry curvature is introduced and non-trivial band topology is induced. Such scheme was tested in Cu[1,3-benzenedicarboxylate(bdc)] by the measurements of the magnon dispersions using inelastic neutron scatterings\cite{CHFSBNL_PRL2015}. Magnon Hall effect\cite{KNL_PRL2010}, as a consequence of the non-zero Berry curvature of the magnon bands, was also observed in such ferromagnetic materials in recent heat transport experiments\cite{OIKSNT_S2010,HCLO_PRL2015}. Other theoretical proposals and experimental realizations include Weyl magnons in ferromagnetic and anti-ferromagnetic pyrochlores\cite{LLKBYC_NC2016,MHJM_PRL_2016,SWW_PRB2017}, topological magnons in honeycomb ferromagnet\cite{O_JPCM2016}, Dirac and nodal line magnons in $\text{Cu}_3\text{TeO}_6$\cite{LLHLF_PRL2017,YLWXDIKLFL_arXiv2017,BWWCLMWRDAWYJW_arXiv2017}, etc.

\par These previous proposals share common features: they are based on local spin models where the linear spin wave theory provides a comprehensive understanding and the DM interaction is crucial to generate non-trivial band topology. Local spin magnetism originates from the exchange interactions between local electron spins. There exists another class of magnetism, the itinerant magnetism, which results from different mechanisms. Especially, due to the lack of exact one local electron spin per physical site in such magnets, the standard linear spin wave theory fails. Then, a natural question of fundamental interest follows: can topological magnons emerge in itinerant magnets? If so, what is the mechanism that leads to itinerant topological magnons? Here, we will investigate this with a one-dimensional interacting electronic model with a flat band as the prototype.

\par The existence of flat electron bands is one of the simplest mechanisms that result in itinerant ferromagnetism\cite{T_PRL1992,M_PLA1993,MT_CMP1993}. The stability of the ferromagnetic ground state and low energy spin wave excitations in such systems have been investigated in several works\cite{T_PRL1994,KA_PRL1994,DG_PRB2015}, yet the band topology of such magnons has long been ignored. Detailed studies of the topological properties of magnons in itinerant magnets are still lacking and highly deserved. In this paper, we consider a one-dimensional Tasaki model\cite{T_PRL1992,T_PTP1998}, which can be viewed as a quarter filled periodic Anderson model with impurities located at the center of bonds of two nearest-neighbor lattice sites and hybridizing with itinerant electrons at these sites, together with a Hubbard repulsion for itinerant electrons. When the on-site energy of impurities is properly tuned, an exact flat electron band emerges and is separated from an upper band by a gap. Then the on-site Hubbard interactions lead the ground state to be a ferromagnetic insulator. The gap between the flat band and upper band is taken to be the largest energy scale in this paper. Therefore, one can project the Hubbard interactions onto the half-filled flat band\cite{KA_PRL1994,T_PTP1998}, and in this way a much larger system than the usual exact diagonalization (UED) method can be numerically accessed. The spin-1 excitation spectrum is calculated by this projected exact diagonalization (PED) method and the low-energy part exhibits well defined magnon band structures, with an acoustic and an optical magnon band. By increasing the Hubbard interaction between conducting electrons, the gap between the acoustic and optical magnons closes and reopens while the Berry phase of the acoustic band changes from 0 to $\pi$, suggesting the occurrence of a topological transition. After this transition, consistent with the bulk-edge correspondence, there always exist in-gap edge magnonic modes. The topological magnons are shown to be stable against several perturbations, such as the non-flatness of the lower electron band, the nearest neighbor interactions between electrons, etc. Therefore, we elaborate the existence of topological magnons in itinerant electron ferromagnets.

\par The rest of the paper is organized as follows. In Sec.\ref{mm}, we introduce our prototype model and formulate the PED method. In Sec.\ref{nr}, we summarize the phase diagram of the model by presenting the numerical results of the magnon bands, Berry phase and edge states. Sec.\ref{stm} discusses the stability of the topological magnons against perturbations. Sec.\ref{sd} provides a summary and discussion.

\section{Model and method}\label{mm}
\subsection{Introduction of model}
\par
\begin{figure}
\includegraphics[scale=0.4]{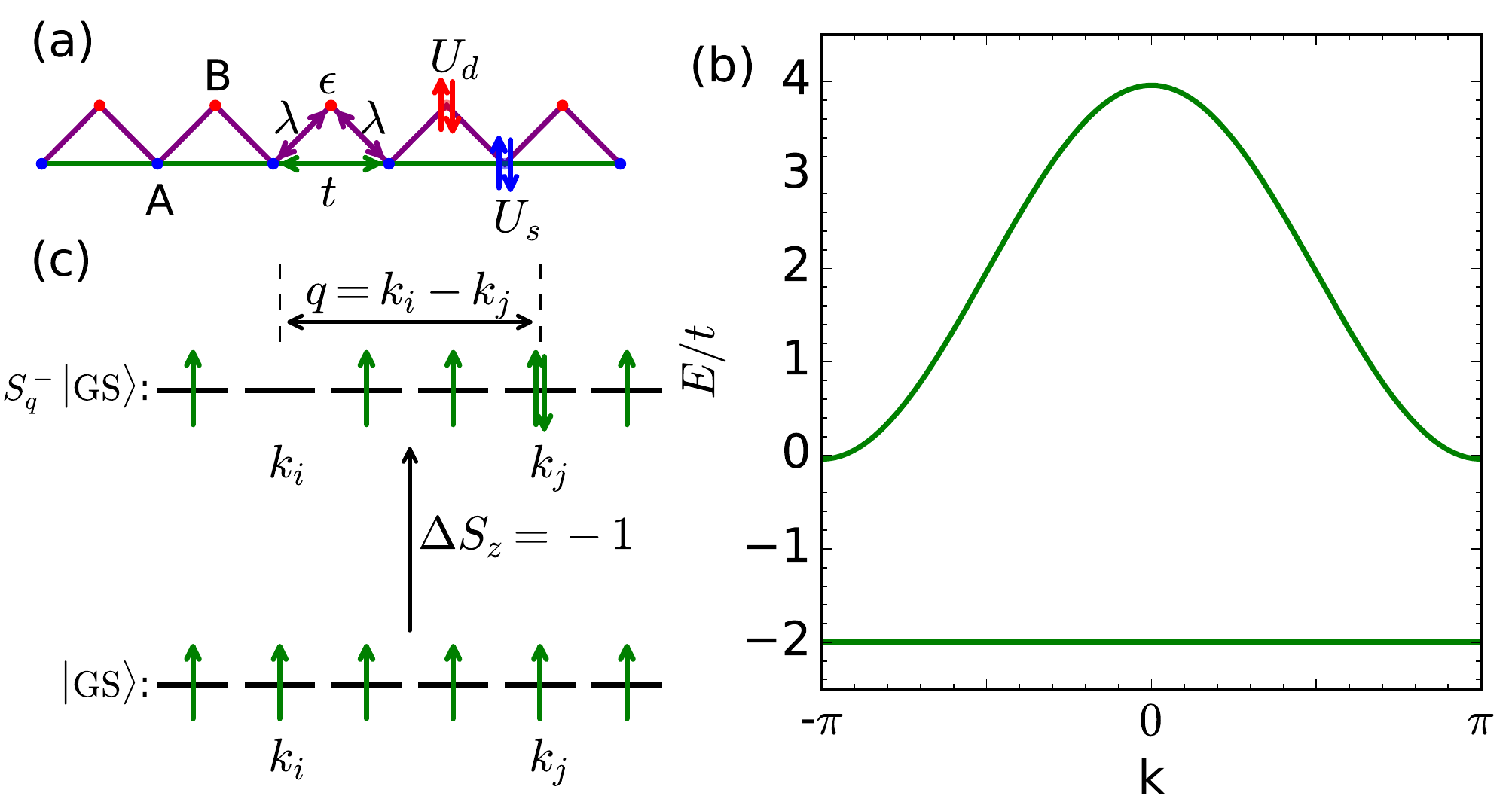}
\caption{(color online). (a) Schematic illustration of Eq.\ref{model}. A and B denote the two inequivalent sites. The hopping amplitude $t$, hybridization $\lambda$, on-site energy $\epsilon$, and Hubbard interactions $U_s$ and $U_d$ are also shown. (b) Band structure of the free part of Eq.\ref{model} with $\epsilon=\lambda^2/t-2t$ which leads to an exact flat band. Here, $\lambda$ is taken to be $1.4t$. (c) Illustration of the creation of a spin-1 excitation with a center-of-mass momentum $q$ on the ferromagentic ground state in the PED method.}
\label{lattice}
\end{figure}

 The electron model we consider, as is illustrated in Fig.\ref{lattice}(a), can be written as
\begin{equation}\label{model}
\begin{aligned}
    H   &   =t\sum_{\langle ij\rangle_{AA}}c_{iA}^\dagger c_{jA}+h.c.+ \lambda\sum_{\langle ij\rangle_{AB}}c_{iA}^\dagger c_{jB}+h.c.\\
        &   +\epsilon\sum_i c^\dagger_{iB}c_{iB}+U_s\sum_i n_{iA\uparrow}n_{iA\downarrow}+U_d\sum_in_{iB\uparrow}n_{iB\downarrow}
\end{aligned}
\end{equation}
Here, $\langle ij\rangle_{AA}$ denotes the nearest-neighbor bonds among A sites and $\langle ij\rangle_{AB}$ denotes the nearest-neighbor bonds between A and B sites, $\epsilon$ is the onsite energy at B sites. $U_s$ and $U_d$ are the Hubbard repulsions at A and B sites, respectively. Others are in standard notation. This model can be viewed as a modified periodic Anderson model, with A sites hosting the conducting electrons and B sites acting as impurities located at the center of the nearest-neighbor AA bonds. A standard periodic Anderson model ignores the $U_s$ term considering that it is usually small compared to $U_d$ at half-filling. Here, $U_s$ is essential to the generation of ferromagentism in the quarter-filled case\cite{T_PRL1992}.

\par When $\epsilon=\lambda^2/t-2t$, the free part of the model possesses an exact flat electron band, which is separated from the upper band by a gap equal to $\lambda^2/t$, as shown in Fig.\ref{lattice}(b). We will first discuss the results related to the flat-band in the lowest energy, and then discuss the effects of a dispersive lowest energy band in Sec.\ref{stm}. Here, $\lambda$ is taken to be larger than both $U_s$ and $U_d$. Therefore the Hubbard interactions ($U_s$ and $U_d$ terms) can be projected onto the flat band. As is well known, the ground state of a half-filled flat electron band with Hubbard interactions is an itinerant ferromagnet. Then, the creation of a spin-1 excitation with a center-of-mass momentum $q$ from the fully spin-polarized ground state is just to choose a single particle state with momentum $k_i$, flip the electron's spin on it and move it onto another single particle state with momentum $k_j=k_i-q$, as illustrated in Fig.\ref{lattice}(c). Thus, the total dimension of the Hilbert space of spin-1 excitations with a definite center-of-mass momentum is linear with respect to the system size $N_q$, which is in sharp contrast to the exponential dependence in the UED method. Therefore, a much larger system can be numerically accessed by this PED method, as presented below we can obtain the numerical results with lattice sites $N_q$ up to $800$(800 A sites and 800 B sites).

\subsection{Formulation of PED method}
\par To perform the projection of the interactions between electrons onto the flat band, the free part $H_0$ of the Hamiltonian must be diagonalized first. Considering the $z$ component of the electron spin is a conserved quantity, as is the case of Eq.\ref{model} in this paper, we can write $H_0$ in momentum space in a general form as follows
\begin{equation}\label{H0}
H_0=\sum_k
\begin{bmatrix}
    c^\dagger_{k\uparrow} & c^\dagger_{k\downarrow}
\end{bmatrix}
\begin{bmatrix}
    H^{\uparrow}(k) & 0 \\
    0 & H^{\downarrow}(k)
\end{bmatrix}
\begin{bmatrix}
    c_{k\uparrow} \\
    c_{k\downarrow}
\end{bmatrix}
\end{equation}
Here, $c_{k\sigma}=\left[c_{k\sigma 1},\cdots,c_{k\sigma m}\right]$ is a $1\times m$ vector, $H^{\sigma}$ is a $m\times m$ matrix, $\sigma=\uparrow\downarrow$, and $m$ denotes the number of orbital indices (the site index is considered as a generalized orbital index) of the electron operator in a unit cell. Take a linear superposition of the $c$ operator, $\alpha_{k\sigma\mu}=\sum_{i=1}^m\left[\mathcal{U}^\sigma(k)\right]_{\mu i}^\dagger c_{k\sigma i}$, which is used to diagonalize $H_0$, i.e. $\left[\mathcal{U}^\sigma(k)\right]^\dagger_{\mu'i'}H^\sigma_{i'i}(k)\mathcal{U}^\sigma_{i\mu}(k)=\varepsilon_{\sigma\mu}(k)\delta_{\mu',\mu}$, here $\mu$ denotes the band index, $i$ the orbital index, and $\varepsilon$ the eigen single-particle energy. Let $\mathcal{P}$ be the set of the indices of the bands onto which we want to project the original Hamiltonian, and $P$ the corresponding projector, then
\begin{equation}\label{projection}
  P^\dagger c_{k\sigma i}P=\sum_{\mu\in\mathcal{P}}\mathcal{U}^\sigma_{i\mu}(k)\alpha_{k\sigma\mu}
\end{equation}
For the model Eq.\ref{model} we considered in this paper, $\mathcal{P}$ contains nothing but the flat band. However, for a general multi-band system, $\mathcal{P}$ should contain all the indices of the bands that contribute to low-energy physics. We keep the summation over $\mathcal{P}$ in Eq.\ref{projection} explicitly because it can deal with our model Eq.\ref{model} subject to open boundary conditions as well by just ignoring the $k$ index. The ground state $|\text{GS}\rangle$ of a half-filled flat band with Hubbard interactions is a ferromagnet,
\begin{equation}\label{GS}
|\text{GS}\rangle=\prod_{\mu\in\mathcal{P}}\prod_{k}\alpha^\dagger_{k\uparrow\mu}|\text{Vac}\rangle
\end{equation}
Here, $|\text{Vac}\rangle$ denotes the vacuum of electrons. After the projection, as shown in Fig.\ref{lattice}(c), the basis for a spin-1 excitation with a center-of-mass momentum $q$ is
\begin{equation}\label{basis}
  |q,k,\mu\mu'\rangle=\alpha^\dagger_{k-q\downarrow\mu}\alpha_{k\uparrow\mu'}|\text{GS}\rangle
\end{equation}
Here we have included the band indices $\mu$ and $\mu'$ for a general consideration.
It is straightforward to get the operation of $H_0$ on the above basis
\begin{equation}\label{PH0P}
\begin{aligned}
  &\langle q',k',\nu\nu'|'P^\dagger H_0P|q,k,\mu\mu'\rangle \\
  =& \left[\varepsilon_{\downarrow\mu}(k-q)-\varepsilon_{\uparrow\mu'}(k)\right]\delta_{\mu,\nu}\delta_{\mu',\nu'}\delta_{k,k'}\delta_{q,q'}
\end{aligned}
\end{equation}
As to the Hubbard interactions $H_U$, which in momentum space can be written as
\begin{equation}\label{Hubbard}
  H_U=\frac{1}{N}\sum_{i}U_i\sum_{kk'p}c^\dagger_{k-p\uparrow i}c_{k\uparrow i}c^\dagger_{k'+p\downarrow i}c_{k'\downarrow i}
\end{equation}
After some algebra, we have
\begin{widetext}
\begin{equation}\label{PHUP}
\begin{aligned}
  &P^\dagger H_U P|q,k,\mu\mu'\rangle    \\
    =&\frac{1}{N}\sum_iU_i\left[
        \sum_\nu\sum_{k'\neq k\;\text{or}\;\nu'\neq\mu'}\mathcal{U}^{\uparrow*}_{i\nu'}(k')\mathcal{U}^\uparrow_{i\nu'}(k')
        \mathcal{U}^\downarrow_{i\mu}(k-q)\mathcal{U}^{\downarrow*}_{i\nu}(k-q)|q,k,\nu\mu'\rangle
    \right]     \\
    -&\frac{1}{N}\sum_iU_i\left[
        \sum_\nu\sum_{\nu'\neq\mu'}\mathcal{U}^{\uparrow*}_{i\mu'}(k)\mathcal{U}^\uparrow_{i\nu'}(k)
        \mathcal{U}^\downarrow_{i\mu}(k-q)\mathcal{U}^{\downarrow*}_{i\nu}(k-q)|q,k,\nu\nu'\rangle
    \right]     \\
    -&\frac{1}{N}\sum_iU_i\left[
        \sum_{p\neq0}\sum_{\nu,\nu'}\mathcal{U}^{\uparrow*}_{i\mu'}(k)\mathcal{U}^\uparrow_{i\nu'}(k+p)
        \mathcal{U}^\downarrow_{i\mu}(k-q)\mathcal{U}^{\downarrow*}_{i\nu}(k+p-q)|q,k+p,\nu\nu'\rangle
    \right]
\end{aligned}
\end{equation}
\end{widetext}
Equipped with Eq.\ref{PH0P} and Eq.\ref{PHUP}, we can obtain the quantities we computed in this paper.

\section{Numerical Results}\label{nr}
\begin{figure}
\includegraphics[scale=0.41]{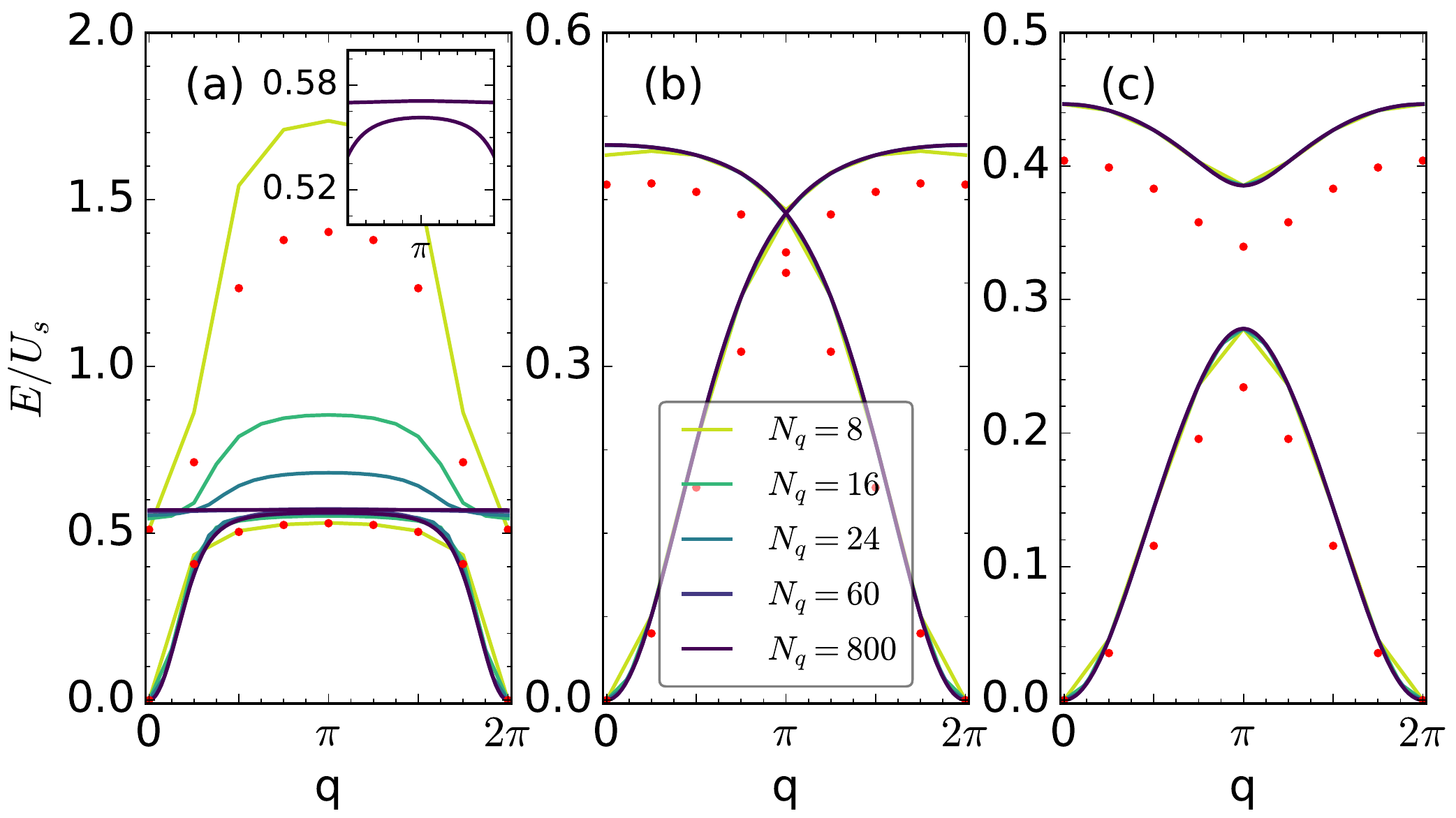}
\caption{(color online). Mangon bands calculated by PED (solid lines) and UED (red dots, $N_q=8$) with (a) $U_s=0.02$, (b) $U_s=0.40$, (c) $U_s$=0.78. Other parameters are fixed at $t=1.0$, $\lambda=1.4$, $U_d=1.0$. The inset in (a) shows the magnon bands calculated by PED with $N_q=800$ around $q=\pi$.}
\label{spectrum}
\end{figure}

\par The low energy spectra of spin-1 excitations out of the flat-band ferromagnetic state obtained by the PED method are shown in Fig.\ref{spectrum} as solid lines for several $U_{s}$, together with the results by the UED method as red dots for a comparison. The UED results are calculated with $N_q=8$ which is the maximum size we can access, and the PED results with $N_q$ ranging from as small as $8$ to as large as $800$. One can see that the PED results converge when $N_q\ge 60$ for all $U_s$ considered here, though the size dependence is obvious when $N_q< 60$ in the case of a small $U_s=0.02$ as shown in Fig.\ref{spectrum}(a). At the same time, the PED and UED results agree with each other qualitatively, verifying that the projection onto the flat-band is applicable. Therefore, we will start our discussion based on the results calculated with the PED method with $N_q=800$ in the following, which are shown as purple lines in Fig.\ref{spectrum}. One can see that the low energy spectra exhibit well defined band structures. As is common in local spin models, we identify the lower band as the acoustic magnon band and the upper band as the optical magnon band. Above these well defined collective modes, we have observed the high energy Stoner continuum, which is not shown here\cite{KA_PRL1994}. At $q=0$, the acoustic magnon band is gapless, which is the character of a ferromagnetic excitation with the Goldstone mode. For a small $U_s$, there exists a gap between the acoustic and nearly flat optical magnon band, which can be seen more clearly from the inset of Fig.\ref{spectrum}(a). Interestingly, the gap closes at $q=\pi$ when $U_s$ is increased to be near 0.409, and reopens with the further increase of $U_s$, suggesting a kind of transition [see Fig.\ref{spectrum}(b) and (c)]. However, from the point of view of PED, the ground state of Eq.\ref{model} is nothing but the Fermi sea filling all the spin-up single particle states of the flat electron band, which is merely determined by the free part of Eq.\ref{model}. Thus, the ground state remains unchanged during this transition and no local order parameter can be used to distinguish the two phases separated by this transition. Therefore, this transition is not a traditional phase transition.

\par
\begin{figure}
\includegraphics[scale=0.35]{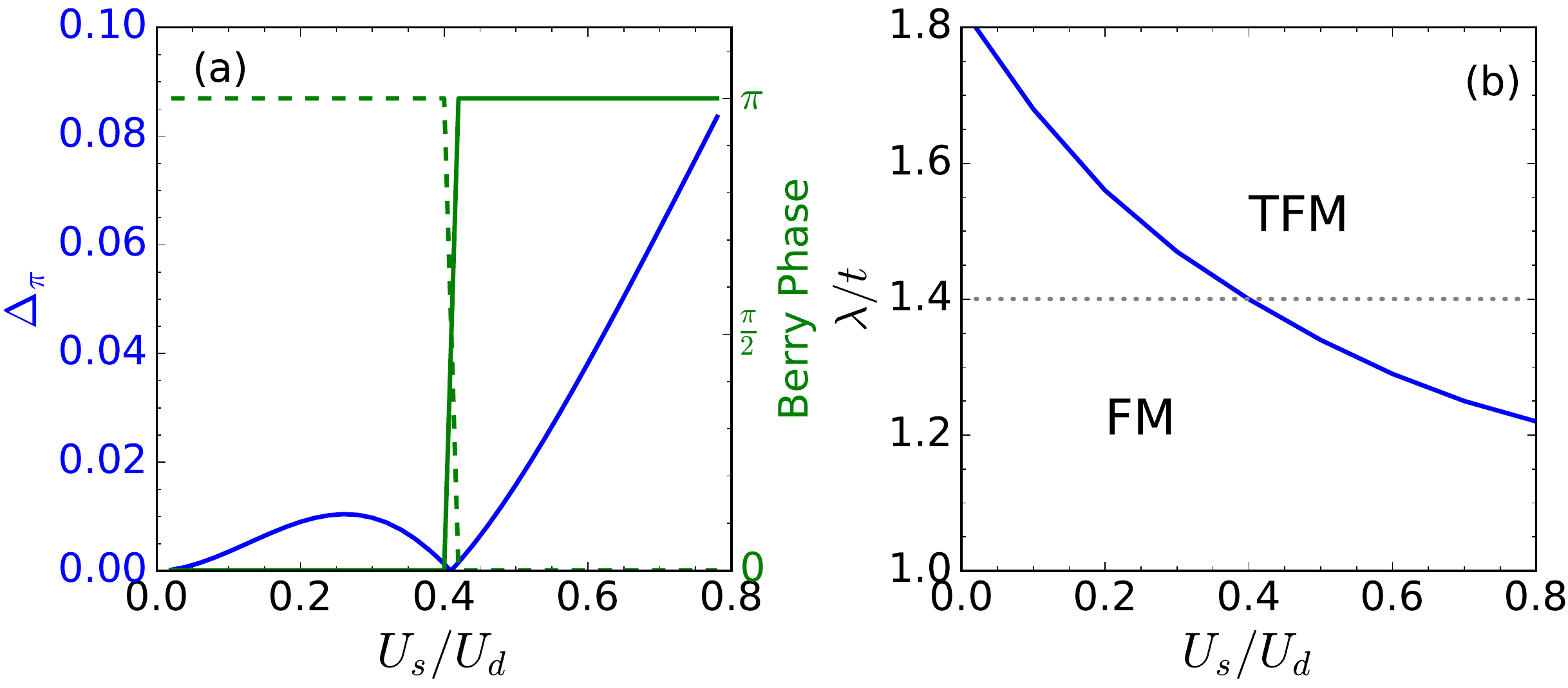}
\caption{(color online). (a) Quantities obtained by the PED method to determine the phase digram shown in Fig.\ref{gap}(b). The blue line represents the magnitude of the gap between the acoustic band and optical band at $q=\pi$ with $N_q=800$. The green solid line and green dashed line represent the corresponding Berry phase of the acoustic band and optical band with the same parameters, respectively. Other parameters are fixed at $t=1.0$, $\lambda=1.4$, $U_d=1.0$. (b) Phase diagram of Eq.(\ref{model}). FM and TFM denotes ferromagnetic  and topological ferromagnetic acoustic magnons, respectively. The dotted line marks the parameters used in Fig.\ref{gap}(a) and in Fig.\ref{edge}(a).}
\label{gap}
\end{figure}

 To explore the nature of the transition with $U_s$, we resort to the study of the band topology. The generic topological information of a one-dimensional band is encoded in its Berry phase, which can be represented as
\begin{equation}
\gamma=\text{Im}\oint\langle \Psi_q|\frac{\partial}{\partial q}|\Psi_q\rangle dq
\end{equation}
where $|\Psi_q\rangle$ denotes the corresponding eigenstate of the band with momentum $q$. The calculation of $\gamma$ requires the comparison between eigenstates at different points in momentum space, which in general, belong to different Hilbert spaces. Mathematically, such a comparison needs the parallel transport of eigenstates from one point to another in momentum space. While physically, it can be resolved by a natural indexing of the basis of the Hilbert spaces at different momentum points. Our PED method provides such a natural indexing. As illustrated in Fig.\ref{lattice}(c), the basis for spin-1 excitations with center-of-mass momentum $q$ can be represented as $|q,k_i\rangle=\alpha_{k_i-q\downarrow}^\dagger\alpha_{k_i\uparrow}|\text{GS}\rangle$, where $\alpha_{k_i-q\downarrow}^\dagger$ creates a spin-down electron on the flat band with momentum $k_i-q$ and $\alpha_{k_i\uparrow}$ annihilates a spin-up electron on the flat band with momentum $k_i$. Thus, the two-particle basis can be viewed as a multi-orbital single particle basis, with $k_i$ acting as the orbital index. Then, analogous to the systems of free particles, the calculation of the Berry phase for itinerant magnon bands becomes straightforward. In Fig.\ref{gap}(a), we show the Berry phase $\gamma$ as a function of the ratio $U_s/U_d$, together with the gap between the acoustic and optical magnon bands $\Delta_\pi$ at $q=\pi$. The green solid line and green dashed line represent the Berry phase of the acoustic and optical magnon band, respectively, and the blue solid line represents $\Delta_\pi$. One can find that the optical magnon band has already acquired a $\pi$  Berry phase for small $U_s$ values, but the acoustic magnon band exhibits the topological trivial property with a zero Berry phase. Remarkably, accompanying the closing and reopening of the gap $\Delta_\pi$, the Berry phase of the acoustic magnon band changes from $0$ to $\pi$, while that of the optical band changes from $\pi$ to $0$. We thus identify the Hubbard interaction $U_s$ driven transition as a topological transition, after which the acoustic magnon band becomes topological nontrivial.

\par The above discussions can be summarized by the phase diagram shown in Fig.\ref{gap}(b). Here, FM denotes ferromagnetic acoustic magnons and TFM denotes topological acoustic magnons. An interesting fact about the phase diagram is, for a fixed ratio of $U_s/U_d$, the increase of the hybridization $\lambda$ between conduction electrons and local impurities can also drive the system from the FM phase to TFM phase. This implies that $\lambda$ and $U_s$ play a cooperative role in the generation of topological acoustic magnons. This is because the onsite energy of the B sites $\epsilon=\lambda^2/t-2t$ also increases with the increase of $\lambda$, which results in transfers of electrons from the B site to A site, and thus enhances the effect of the Hubbard interactions on the A site.

\begin{figure}
\includegraphics[scale=0.48]{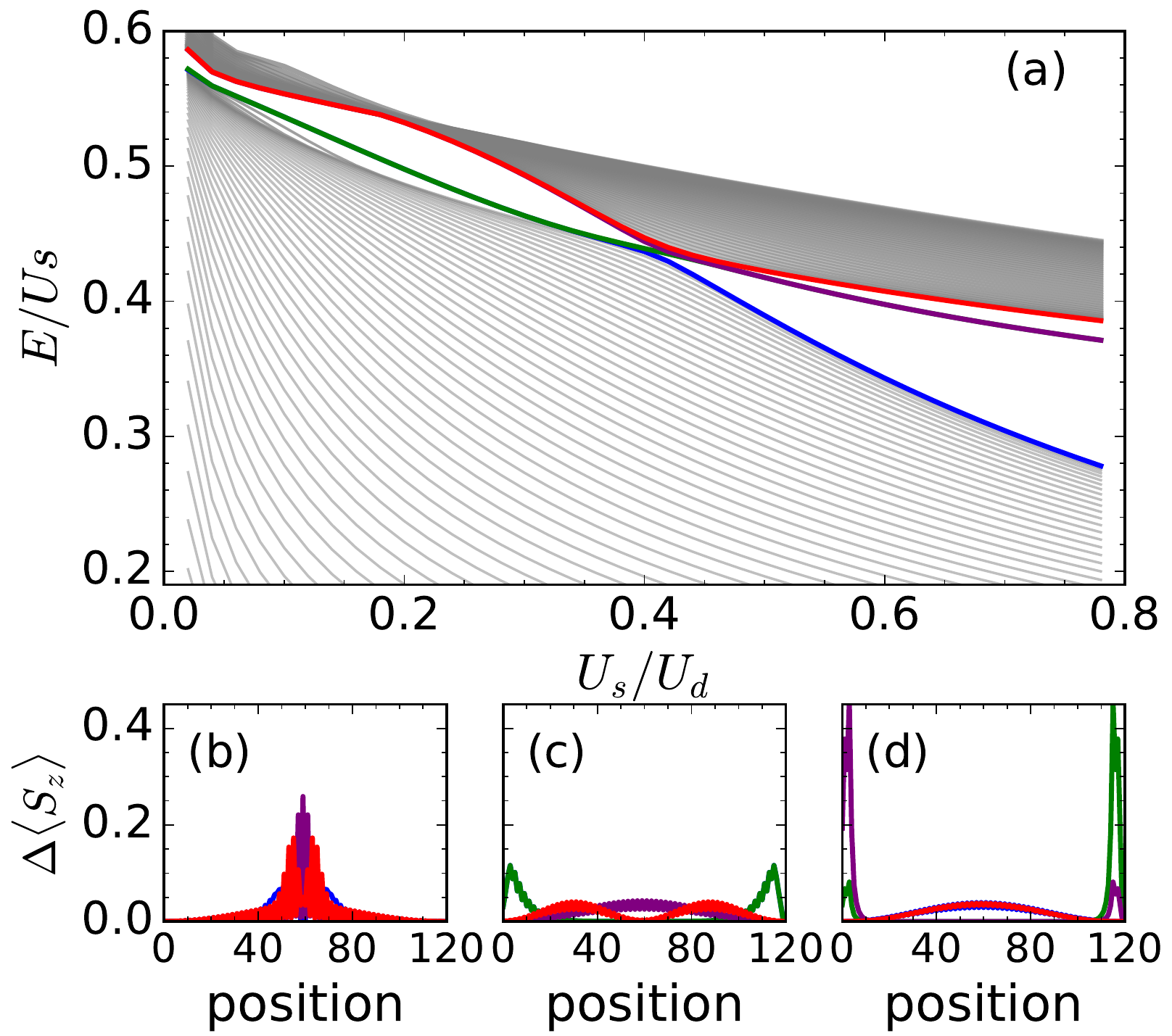}
\caption{(color online). At the top: (a) Magnon spectrum of Eq.\ref{model} subject to open boundary conditions. At the bottom: Difference of the $S_z$ profile between the in-gap or near-gap magnon states and ground state with (b) $U_s=0.02$, (c) $U_s=0.30$ and (d) $U_s=0.78$. Other parameters are fixed at $t=1.0$, $\lambda=1.4$, $U_d=1.0$.}
\label{edge}
\end{figure}

\par According to the bulk-edge correspondence, the acoustic magnon band in the TFM phase, which hosts a non-zero Berry phase, should lead to the localized in-gap magnonic modes when the system is subject to open boundary conditions. So, let us now check the existence of the edge states for magnonic excitations. We note that the exact definition of edge states in a many body system is rather subtle\cite{LGLW_NJP2017}. Here, we use a simple but sensible method to detect edge magnonic modes: we compute the expectation value of the $z$ component of electron spins both on the ground state and a relevant state, then use their difference, i.e., the quantity
\begin{equation}
\Delta\langle S_z\rangle=\langle\text{M}|S_z|\text{M}\rangle-\langle\text{GS}|S_z|\text{GS}\rangle
\end{equation}
as a measurement of the edge magnonic modes, where $|\text{M}\rangle$ denotes a state used to test if it has edge magnonic modes. In Fig.\ref{edge}, we plot the magnon spectrum of Eq.\ref{model} on a geometry with open boundaries, and the profiles of $\Delta\langle S_z\rangle$ for four relevant in-gap or near-gap states which are denoted by red, green, purple and blue lines. In such a system, the dimension of the irreducible Hilbert space of spin-1 excitations scales quadratically with the system size, due to the lack of the center-of-mass momentum as a good quantum number. Therefore, only a restricted system size can be numerically accessed. In Fig.\ref{edge}, the system size is $120$ ($60$ unit cells containing 2 inequivalent A and B sites). As can be seen in Fig.\ref{edge}(a), after the topological transition, there always exist two degenerate in-gap magnonic states which are denoted by the purple and green lines (converge due to degeneracy). In this $U_s$ regime, the $\Delta\langle Sz\rangle$ profile exhibits clear characteristic of edge modes, with one denoted by the purple line on one end and the other by the green line on the other end, as shown in Fig.\ref{edge}(d) for $U_s=0.78$. It is noted that the spectral weight of the edge modes does not concentrate on only one end, instead it leaves a small part of weight on the other end. It arises because the two in-gap magnonic states are exactly degenerate and any superposition of them is also an eigenstate of the system, which may deviate from the ground state at both ends of the chain, so this does not imply a kind of fractionalization of the corresponding magnons. Combining the results of the $\pi$ Berry phase and edge modes after the $U_{s}$ driven transition, we can conclude that there are topological acoustic magnons after $U_{s}\approx 0.409$.

\par We would like to mention that there also seem to be in-gap magnonic edge modes (green line and blue line, degenerate) before the topological transition, as shown in Fig.\ref{edge}(a). In fact, when $U_s/U_d$ is larger than $0.1$, the profile of $\Delta\langle S_z\rangle$ suggests that they are also edge modes as shown in Fig.\ref{edge}(c) for $U_s=0.3$. Yet without the close of the magnon gap, these states are continuously connected to those with $U_s/U_d$ around $0.02$, which do not show edge modes [see Fig.\ref{edge}(b)]. Of course, when $U_s/U_d$ is as small as $0.02$, it can be seen from Fig.\ref{spectrum}(a) and Fig.\ref{gap} that the magnon gap is also very small and the system may suffer from the finite-size effect, and it is unclear whether a chain with $120$ sites is large enough to determine the properties in the thermal dynamic limit. On the other hand, we want to remark that the $\pi$ Berry phase of the optical magnon band before the topological transition does not guarantee the existence of magnonic edge modes in the gap between the acoustic band and optical band according to the bulk-edge correspondence. A possible consequence of this $\pi$ Berry phase is the existence of edge modes in the gap between the optical band and Stoner continuum. However, due to the flatness of the optical band and Stoner continuum [as can be seen in Fig.\ref{spectrum}(a), Fig.\ref{dspectrum}(a) and Fig.\ref{vspectrum}(a)], the relevant eigenstates are highly near-degenerate, which makes its verification rather difficult in the framework of PED method.

\section{Stability of topological magnons}\label{stm}
\par Topological magnons are expected to be stable against perturbations that do not destroy the magnetic order of the ground state or that do not close the magnonic gap between the acoustic band and optical band. In this section, two kinds of perturbations, i.e. the non-flatness of the lower electron band and the nearest neighbor interactions between electrons will be discussed to show the stability of the itinerant topological magnons.

\subsection{Non-flatness of lower electron band}
\par As stated in Sec.\ref{mm}, an exact flat electron band exists when $\epsilon=\lambda^2/t-2t$. When $\epsilon$ deviates this value, the lower electron band will disperse. Let $H_\Delta$ denotes this deviation:
\begin{equation}\label{delta}
H_{\Delta}=\Delta\sum_ic_{iB}^\dagger c_{iB}
\end{equation}

\begin{figure}
\includegraphics[scale=0.46]{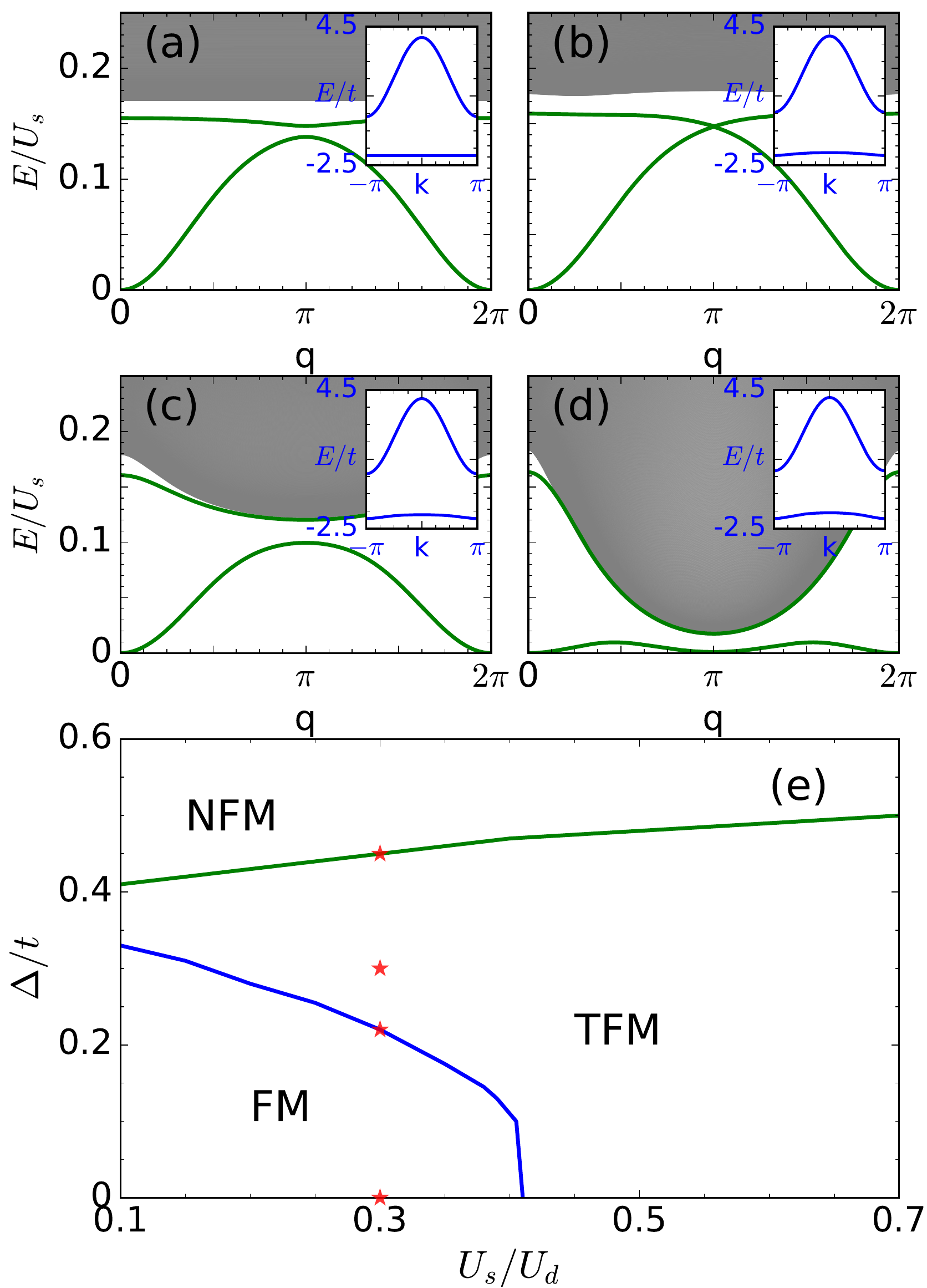}
\caption{(color online). (a)-(d) Spin-1 excitation spectra calculated by PED with $N_q=800$ of Eq.\ref{model} perturbed by Eq.\ref{delta} with (a) $U_s=0.30$, $\Delta=0$, (b) $U_s=0.30$, $\Delta=0.22$, (c) $U_s=0.30$, $\Delta=0.30$, (d) $U_s=0.30$, $\Delta=0.45$. The shaded area represents the Stoner continuum. Insets of (a)-(d) show the corresponding perturbed free electron bands. (e) The phase diagram in the $\Delta$-$U_s$ parameter space. NFM, FM and TFM represent non-ferromagnetic phase, ferromagnetic magnons and topological ferromagnetic magnons, respectively. Red stars mark the parameters used in (a)-(d). Other parameters are fixed at $t=1.0$, $\lambda=1.4$, $U_d=1.0$.}
\label{dspectrum}
\end{figure}

\par The spin-1 excitation spectra of Eq.\ref{model} with $H_\Delta$ are shown in Fig.\ref{dspectrum}(a)-Fig.\ref{dspectrum}(d). The perturbed free electron bands are also shown in the corresponding insets. With $\Delta$, the flat electron band becomes dispersive. As $\Delta$ grows, the gap between the acoustic band and optical band closes [Fig.\ref{dspectrum}(b)] and reopens [Fig.\ref{dspectrum}(c)]. Meanwhile, we find that the Berry phase of the acoustic band changes from $0$ to $\pi$ while that of the optical band changes from $\pi$ to $0$. With the further increase of $\Delta$, the acoustic band touches zero energy at $q=\pi$ [Fig.\ref{dspectrum}(d)], which marks the destablization of the ferromagnetic ground state. This instability results from the competition between the kinetic energy and the potential energy of the electrons, as a fully spin-polarized state minimizes the energy of Hubbard interactions but cost more energy when the electron band disperses.

\par The phase diagram of Eq.\ref{model} with $H_\Delta$ in the $\Delta$-$U_s$ plane are shown in Fig.\ref{dspectrum}(e). Here, NFM, FM and TFM represent non-ferromagnetic phase, ferromagnetic magnons and topological ferromagnetic magnons, respectively. Apparently, TFM survives quite a wide range of parameters as long as the perturbation $H_\Delta$ does not destroy the ferromagnetic ground state. An intriguing fact can also be observed that a small but positive $H_\Delta$ helps to drive the system topologically non-trivial because such a term increases the onsite energy of B sites and thus enhances the Hubbard interaction on A sites, as is similar to the effect of the increase of $\lambda$.

\subsection{Nearest neighbor interaction between electrons}

\begin{figure}
\includegraphics[scale=0.46]{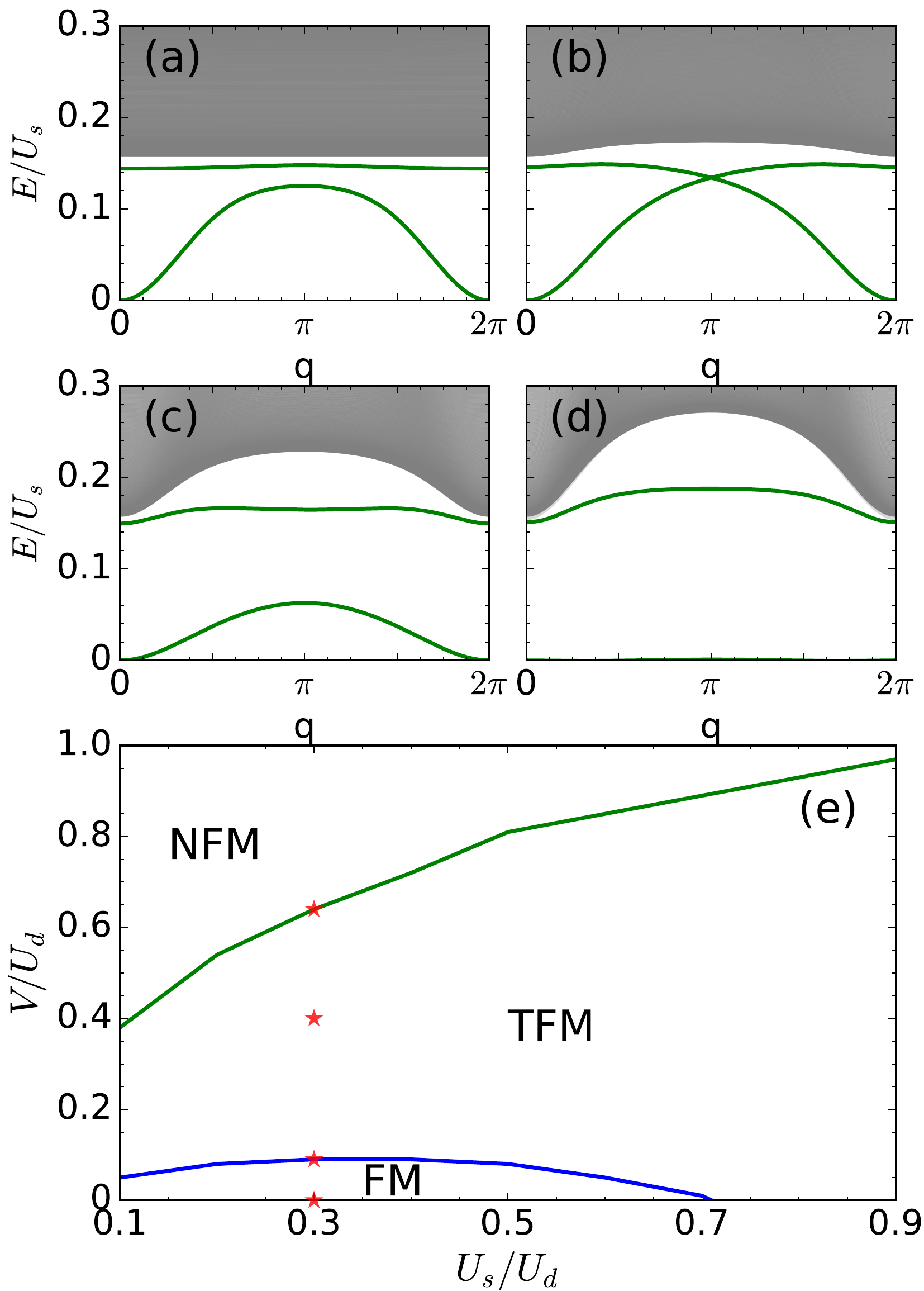}
\caption{(color online). (a)-(d) Spin-1 excitation spectra calculated by PED with $N_q=800$ of Eq.\ref{model} perturbed by Eq.\ref{V} with (a) $U_s=0.30$, $V=0$, (b) $U_s=0.30$, $V=0.09$, (c) $U_s=0.30$, $V=0.40$, (d) $U_s=0.30$, $V=0.64$. The shaded area represents the Stoner continuum. (e) The phase diagram in the $V$-$U_s$ parameter space. NFM, FM and TFM represents non-ferromagnetic phase, ferromagnetic magnons and topological ferromagnetic magnons, respectively. Red stars mark the parameters used in (a)-(d). Other parameters are fixed at $t=1.0$, $\lambda=1.25$, $U_d=1.0$.}
\label{vspectrum}
\end{figure}

\par Another kind of perturbation is the nearest neighbor interactions between electrons $H_V$:
\begin{equation}\label{V}
H_{V}=V\sum_{\langle ij\rangle_{AB}}n_{iA}^\dagger n_{jB}
\end{equation}

\par The spin-1 excitation spectra of Eq.\ref{model} with $H_V$ are shown in Fig.\ref{vspectrum}(a)-Fig.\ref{vspectrum}(d). As is similar to the effect of $\Delta$, with the increase of $V$, the gap between the acoustic band and optical band closes[Fig.\ref{vspectrum}(b)] and reopens[Fig.\ref{vspectrum}(c)], accompanying with the Berry phase of the acoustic band changing from $0$ to $\pi$ and that of the optical band changing from $\pi$ to $0$. Different from previous cases, a larger $V$ strongly suppresses the dispersion of the acoustic band and pushes it to zero[Fig.\ref{vspectrum}(d)]. This instability may result from the competition between charge density wave(CDW) and ferromagnetic order because $H_V$ favors a CDW order. Fig.\ref{vspectrum}(e) is the phase diagram of Eq.\ref{model} with $H_V$ in the $V$-$U_s$ plane. Clearly, TFM also survives in an extended range of the nearest neighbor electron interactions.

\section{Summary and discussion}\label{sd}
\par In summary, we elaborate a new scenario that topological acoustic magnons can emerge from itinerant flat-band ferromagnet, which is different from previous ones that topological magnons exists in local spin models. The prototype model is a one-dimensional flat-band ferromagnet, which can be viewed as a quarter filled periodic Anderson model with impurities located at the center of the bonds and hybridizing with conducting electrons at their neighboring sites. Concentrating on the physics related to the flat-band, we can project the model Hamiltonian onto the flat band and carry out the large-size calculation of spin-1 excitations based on the exact diagonalization method. We find a correlation-driven topological transition to realize non-trivial acoustic magnons, which is driven by the onsite Hubbard repulsion for conducting electrons.

\par The mechanism that results in itinerant topological magnons is different from that in local spin models with DM interactions. On the one hand, DM interactions breaks the spin SU(2) rotation symmetry while our prototype model preserves this symmetry. On the other hand, the particle number per physical site in local spin models is restricted to be 1. Therefore, no charge fluctuation is allowed in such systems and the standard linear spin wave theory applies. Then magnons are understood as the precession of ordered magnetization and the DM interaction acts as a vector potential for the propagation of magnons similar to the magnetic field for electrons. However, no restriction of particle number (other than the Pauli exclusion principle) is a priori required for itinerant magnets, and the low energy spin excitations (acoustic magnons and optical magnons) are accompanied with strong charge fluctuations. Therefore, we can not derive a low-energy effective spin model for itinerant magnets by the usual strong coupling expansion. The lack of such an effective spin model makes it be hard to gain a complete understanding of the emergence of topological magnons, and further works are needed to reveal the microscopic mechanism. Our proposal is based on a one-dimensional model and its possible realization is suggested in cold atom simulations. We also expect to explore other itinerant magnets in higher dimensions.

\begin{acknowledgments}
\par We gratefully acknowledge Zuo-Dong Yu for fruitful discussions. This work was supported by the National Natural Science Foundation of China (11774152) and National Key Projects for Research and Development of China (Grant No. 2016YFA0300401).
\par Xiao-Fei Su and Zhao-Long Gu contribute equally to this work.
\end{acknowledgments}

\bibliography{ref}
\end{document}